\documentclass[fleqn,twoside]{article}
\usepackage{espcrc2}

\usepackage{graphicx,amssymb,subfigure}
\usepackage[figuresright]{rotating}

\newcommand{\AmS}{{\protect\the\textfont2
  A\kern-.1667em\lower.5ex\hbox{M}\kern-.125emS}}

\title{A finite temperature investigation of the Georgi-Glashow
       model in 3D
\thanks{Contribution based on a talk given by A. Barresi.}}

\author{Andrea Barresi\address{Dipartimento di Fisica dell'Universit\`a and INFN, 56127 Pisa, Italy}, 
Luigi Del Debbio\address{CERN, Department of Physics, TH Division, CH-1211 Geneva 23}, 
Biagio Lucini\address{Institute for Theoretical Physics, ETH Z\"urich,
CH-8093 Z\"urich, Switzerland}}
       
\begin{document}

\begin{abstract}
We study the SU(2) gauge theory with scalar matter
in the adjoint representation in 3D at finite temperature.
We find evidence for a finite
temperature phase transition both in the symmetric
and in the broken phase; such transitions are
consistent with the universality class of Ising 2D,
in agreement with recent analytical arguments.
\vspace{1pc}
\end{abstract}

\maketitle

\section{INTRODUCTION}
\label{sec1}
The Georgi-Glashow model in 3D is a useful toy model to investigate
mechanisms of confinement related to topological degrees of
freedom. The model is an SU(2) Yang-Mills theory with a scalar field in the
adjoint representation of the gauge group.
Its Euclidean action in the continuum is written as
{\setlength\arraycolsep{1pt}
\begin{eqnarray}
S=\int & d^3x & \frac{1}{2g^2} \mathrm{Tr} (F_{\mu\nu}F_{\mu\nu}) 
            +  \mathrm{Tr}(D_{\mu}\phi D_{\mu}\phi) \nonumber\\
& + & m^2_0\mathrm{Tr}(\phi \phi)+\frac{\mu}{2}(\mathrm{Tr}(\phi\phi))^2\,,
\end{eqnarray}}
where $D_\mu \phi=\partial_\mu\phi+i[A_\mu,\phi] $ with
 $\phi=\phi^a\sigma^a/2$. The model has 3 dimensionful bare parametrs,
 $g^2=[mass], m_0=[mass],\mu=[mass]$, which can be rewritten in terms
 of 1 dimensionful parameter, $g^2$, and 2 dimensionless,
 $y=m_0^2/g^4, x=\mu/2 g^2$. When the adjoint scalar field acquires a
 non-zero vacuum expectation value (vev), the Higgs mechanism takes place
 and SU(2) is dinamically broken to U(1); as a consequence two gauge
 bosons, $W^\pm$, acquire a mass. At tree level, one finds
 $M_{W^\pm}^2/g^4=-x/y$ and $M^2_\gamma=0$, while the Higgs boson has
 a mass $M_H^2/g^4=2y$ and a vev $\phi^a\phi^a=-y g^2/x$.

Magnetic monopoles solutions~\cite{'tHooft1,Polyakov1} are instantons
in the 3D case.  Monopoles influence both the charged and neutral
sector of the theory: they provide at the same time a mechanism of
confinement of $W^{\pm}$ bosons and a mechanism of mass generation for
the photon field. In the semiclassical approximation one
finds~\cite{Polyakov2}: {\setlength\arraycolsep{1pt}
\begin{eqnarray}
\frac{M_\gamma^2}{g^4}\!\!\sim\!\!\left(\!-\frac{x}{y}\!\right)^{\frac{7}{4}}
\!\!\exp\left(\!\!{-4\pi\sqrt{-\frac{x}{y}}}\right)\,,\,\,
\sigma=\frac{g^2 M_\gamma}{2\pi^2}\,
\end{eqnarray}}
respectively for the photon mass $M_{\gamma}$ and the string tension
$\sigma$.
In our investigation we will use lattice simulations in order to 
understand whether the system undergoes a finite temperature 
deconfinement phase transition and if
the magnetic monopoles are the only relevant degrees of freedom necessary
to describe correctly the critical behaviour of the system near the
critical temperature $T_c$.

\section{THE LATTICE MODEL}
\label{sec2}
The discretized action of the Georgi-Glashow model can be written as
{\setlength\arraycolsep{1pt}
\begin{eqnarray}
S & = & \beta\!\!\sum_{x,\mu>\nu}\!\!
        \Big(\!1-\frac{1}{2}\mathrm{Tr} U_{\mu\nu}(x)\!\Big) 
        \!+\! 2\!\sum_{x}\mathrm{Tr}(\Phi(x)\Phi(x)) \nonumber\\
  & - & 2\kappa\sum_{x,\mu}\mathrm{Tr}(\Phi(x)U_{\mu}(x)\Phi(x+\hat{\mu}a)
        U_{\mu}^{\dagger}(x)) \nonumber\\
  & + & \lambda\sum_{x}(2\mathrm{Tr}(\Phi(x)\Phi(x))-1)^2 \,,
\end{eqnarray}}
where $U_{\mu\nu}(x)$ is the plaquette and the scalar field
$\Phi(x)$ is in the adjoint representation of SU(2). 
In our simulations we used a standard Kennedy-Pendleton heatbath algorithm
for the pure gauge part of the action, modified with a Metropolis step
to take into account the hopping term, which is quadratic in the link.
For the scalar update instead we adopted the Bunk algorithm \cite{Bunk}.
We used lattices with $N_\tau=4,6$ and $N_s=24,32,40,44,48,54,64$.

The model in 3D is superrenormalizable and only a finite number of
counterterms are needed to renormalize the action. The matching of the
lattice theory to the continuum one has been computed in Ref.~
\cite{Laine}. Such equations specify how to approach the continuum
limit along lines of costant physics.

\section{THE PHASE DIAGRAM}
\label{sec3}
\begin{figure}[t]
\begin{center}
\includegraphics[width=0.4\textwidth]{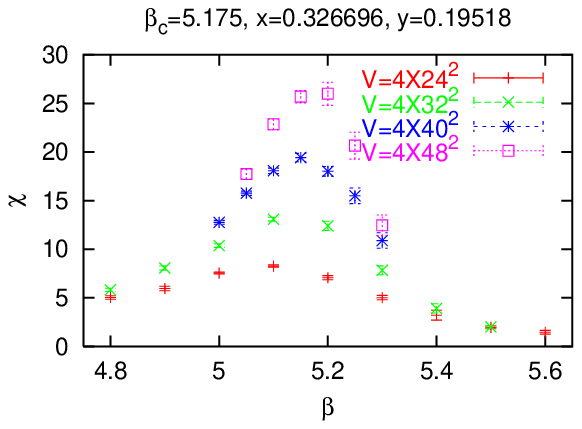}
\includegraphics[width=0.4\textwidth]{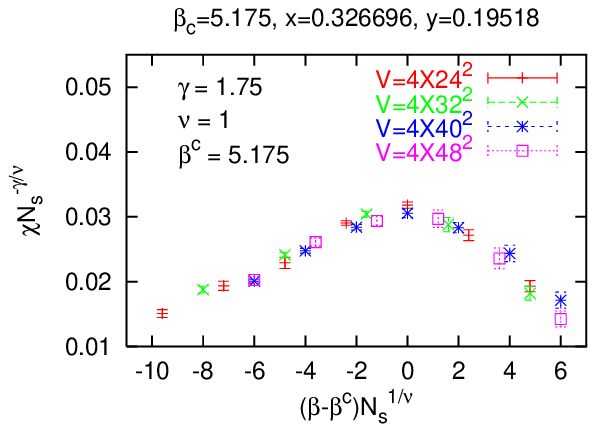}
\end{center}
\vspace{-1cm}
\caption{Susceptibility (up) and fss (down) of $L_F$ at $N_\tau=4$ in the symmetric phase.}
\label{simmT}
\end{figure}

Lattice studies at $T=0$ indicate that the symmetric and Higgs phase
are analytically connected~\cite{Nadkarni,Teper}, i.e. they are
separated by a first order phase transition or crossover depending on
the value of $\lambda$.  What should one expect at finite T?  The
scalar dynamical field is in the adjoint representation, so it cannot
screen static charges in the fundamental representation. Therefore one
should expect that these charges are really confined at zero and low
temperatures and that they undergo a phase transition at a certain
critical temperature.  Moreover the action is $\mathbb{Z}_2$-invariant
and the Polyakov loop can be used to look for the signature of a
finite temperature phase transition. In the symmetric phase we studied
the Polyakov loop in the fundamental representation $L_F$ and its
susceptibility {\setlength\arraycolsep{1pt}
\begin{eqnarray}
\overline{L_F} &=& \frac{1}{N_s^2}\sum_{\vec{x}}\left(\mathrm{Tr}_F
\left(\Pi_{t=1}^{N_\tau}U_\tau\left(\vec{x},t\right)\right)\right)\\
\chi_{L_F} &=& N_s^2\left(\langle\overline{L_F}^2\rangle-
\langle\overline{L_F}\rangle^2\right)
\end{eqnarray}}
across the phase transition, at fixed $N_\tau$ and different spatial
lattice sizes $N_s$. The results are reported in Fig. (\ref{simmT})
(up). The ratio $\chi_{L_F}/N_s^{\gamma/\nu}$ are expected to be a
universal function of the scaling variable, i.e.
\begin{eqnarray}
\chi_{L_F}/N_s^{\gamma/\nu}=f((\beta-\beta_c)N_s^{1/\nu})\,;
\end{eqnarray}
indeed the susceptibilities for different spatial volumes collapse on
the same curve (Fig. (\ref{simmT}) (down)) when rescaled with the
critical indices of the 2D Ising model ($\nu=1,\gamma=1.75$),
suggesting that the Georgi-Glashow model in 3D in the symmetric phase
is in the same universality class of Ising 2D.

Even if the symmetric and the Higgs phase are analitically connected,
the Higgs phase needs particular care both in the analytical and in
the numerical investigation. Following Ref.~\cite{Agasian}, one can
assume that, at low temperatures, monopoles are the only relevant
degrees of freedom. At finite $T$ one dimension is compactified, with
length $\beta=1/T$, and monopoles at distances larger than $\beta$
feel a 2-dimensional interaction; hence the model is described by a
2-dimensional Coulomb potential. A 2-dimensional Coulomb gas undergoes
a BKT phase transition; the conclusion in~\cite{Agasian} is thus that
the model is in the universality class of U(1) 2d.

The analysis presented in~\cite{Dunne}, reaches a different conclusion
by taking into account the effect of the charged $W^{\pm}$
bosons. Dimensional reduction is considered to be a valid
approximation but the starting point is an effective Lagrangian
written in terms of a field $V(x)$, which is the creation operator of
a magnetic vortex with flux $2\pi/g$; this picture allows for a
description of the charged sector of the model. The monopoles play an
essential role in this scenario; indeed the effective theory without
the monopole-induced term would be equivalent to an $XY$ model and
thus it would be again in the $U(1)$ universality class.  Including
the monopole effects, the original $U(1)$ simmetry becomes anomalous
and only the $\mathbb{Z}_2$ subgroup is conserved.  The conclusion in
Ref.~\cite{Dunne} is that the model is in the Ising 2d universality
class.

\begin{figure}[t]
\begin{center}
\includegraphics[width=0.4\textwidth]{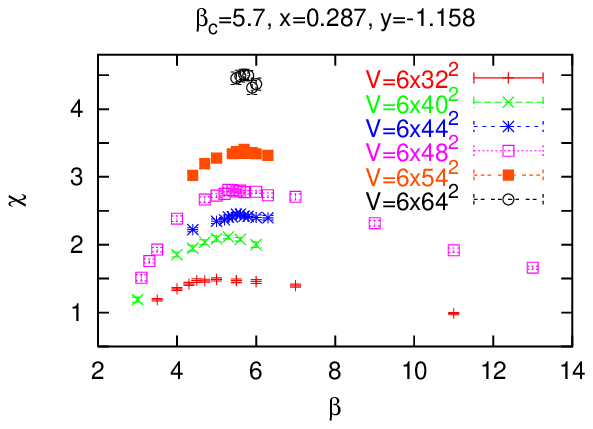}
\includegraphics[width=0.43\textwidth]{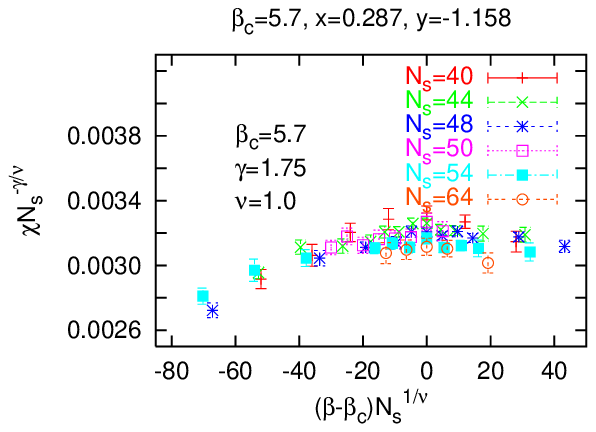}
\end{center}
\vspace{-1cm}
\caption{Susceptibility (up) and fss (down) of $L^m$ at $N_\tau=4$ in the Higgs phase.}
\label{brokT}
\end{figure}

In order to check these two scenarios, we used a modified Polyakov loop \cite{Karsch}
and its susceptibility
{\setlength\arraycolsep{1pt}
\begin{eqnarray}
\overline{L^m} &=&
\frac{1}{N_s^2}\sum_{\vec{x}}\left(\mathrm{Tr}_F\left(\Pi_{t=1}^{N_\tau}U_\tau\left(\vec{x},t\right)\right)\right)^2 \nonumber\\
    &+&\frac{1}{N_s^2}\sum_{\vec{x}}\left(\mathrm{Tr}_F\left(\Phi(\vec{x},t)\Pi_{t=1}^{N_\tau}U_\tau\left(\vec{x},t\right)\right)\right)^2 ,\\
\chi_{L^m}   &=& N_s^2\left(\langle\overline{L^m}^2\rangle-\langle\overline{L^m}\rangle^2\right)\,.
\end{eqnarray}}
In the Higgs phase the gauge links are aligned along the direction of
the scalar field in color space and the Polyakov loops with
$\phi$-insertions give a better signal.  We made simulations deep in
the Higgs phase at couplings which correspond to the perturbative
continuum ratios $M_H/g^2\sim 1.5$ and $M_W/g^2\sim 0.5$.  As one can
see from Fig. (\ref{brokT})(up), the susceptibility increases by
increasing the spatial volume; by rescaling the results at different
volumes with the Ising critical indices, one can see again a nice
collapse on the same curve (Fig. (\ref{brokT})(down)), thus providing
evidence that in the Higgs phase the model is also in the Ising
2D universality class.

\section{CONCLUSIONS}
In our lattice investigation we have found that the Georgi-Glashow model in 3D
undergoes a finite temperature confinement/deconfinement phase transition
characterized by the critical indices of the Ising 2D universality
class, both in the symmetric and in the Higgs phase, suggesting that
at $T\neq 0$ both magnetic monopoles and
vortices must be taken into account in order to get a correct description
of the theory.


\begin{thebibliography}{9}
\bibitem{'tHooft1} G. 't Hooft, Nucl. Phys. B79 (1974) 276.
\bibitem{Polyakov1} A.M. Polyakov, JETP Letters 20 (1974) 194.
\bibitem{Polyakov2} A.M. Polyakov, Nucl. Phys. B120 (1977) 429.
\bibitem{Bunk} B. Bunk, Nucl. Phys. Proc. Suppl.42 (1995) 566. 
\bibitem{Laine} M. Laine, Nucl. Phys. B451 (1995) 484.
\bibitem{Nadkarni} S. Nadkarni, Nucl. Phys. B334 (1990) 559.
\bibitem{Teper} A. Hart, O. Philipsen, J. D. Stack and M. Teper, Phys. Lett. B396 (1997) 217.
\bibitem{Agasian} N.O. Agasian and K. Zarembo, Phys. Rev. D57 (1998) 2475.
\bibitem{Dunne} G.V. Dunne, I.I. Kogan, A. Kovner and B. Tekin, JHEP 01 (2001) 32.
\bibitem{Karsch} F. Karsch, E. Seiler, I.O. Stamatescu, Phys. Lett. B131 (1983) 138.
\end{thebibliography}
\end{document}